# Bibliometric assessment of national scientific journals

Henk F. Moed [1], Felix de Moya-Anegon [2], Vicente Guerrero-Bote[3], Carmen Lopez-Illescas[4], Myroslava Hladchenko[5]

[1] henk.moed@uniroma1.it
Sapienza University of Rome, Italy

[2] felix.moya@scimago.es
SCImago Group, Madrid, Spain

[3] guerrero@unex.es
SCImago Group, Dept. Information and Communication, University of Extremadura, Badajoz, Spain

[4] carmlopz@gmail.com
University Complutense of Madrid. Information Science Faculty. Dept. Information and Library Science, SCImago Group, Spain

[5] hladchenkom@gmail.com
Nicolaus Copernicus University, Toruń, Poland, and National University of Life and Environmental Sciences of Ukraine, Kiev, Ukraine

## Abstract

Nationally oriented scientific-scholarly journals are considered from a methodological-informetric viewpoint, analysing data extracted from Scimago Journal Rank based on Scopus. An operational definition is proposed of a journal's degree of national orientation based on the geographical distribution of its publishing or citing authors, and the role of international collaboration and a country's total publication output. A comprehensive analysis is presented of trends up until 2019 in national orientation and citation impact of national journals entering Scopus, extending outcomes in earlier studies. A method to analyse national journals of given countries is applied to the set of former USSR republics and Eastern and Central European states which were under socialism, distinguishing between *domestic* and *foreign* national journals. The possible influence is highlighted of factors related to a journal's access status, publication language and subject field, international scientific migration and collaboration, database coverage policies, the size of a national research community, historical-political factors and national research assessment and funding policies.

## 1. Introduction

National journals are periodical publications in their own right. They have proper values and functions in the scientific-scholarly communication process. Moreover, they constitute an important subject in library and information science and in scientometrics. National journals are used predominantly by researchers from a particular country to communicate their research results to each other and to an interested audience from that country. To characterize a journal's national orientation a series of features can be considered. The common denominator in most of these is the focus on the geographical location of the various actors: journal publisher; journal editor; manuscript reviewer; publishing author; and reader. Other aspects relate to the relevance broadness of the topics addressed in a journal (local-national-global) or to its publication language.

A study of the use and value of national journals involves a series of actors who all have their own practices, norms and objectives. First of all, *authors* decide on the content and format of their papers, including their publication language, and on where to submit their manuscripts. *Publishers* adopting a certain business model make publications available and determine their access status. *Indexers* review journals and compile indexes to facilitate information searches on a given topic. Indexers of large *citation indexes* such as Clarivate Analytics' Web of Science, Elsevier's Scopus or Google Scholar establish criteria for sources which are processed for their indexes, combining expert knowledge and informetric techniques. *National governments* may stimulate their national *publishing industries*. They may also impose criteria or formula for the *funding* of academic institutions and for the assessment of staff members for *hiring and promotion*. *Research funders* use information extracted from journals and other sources to monitor the success of their funding policies. *Research managers and evaluators* use this information to assess and shape research activities and performance of individual researchers, groups, institutions or national systems. And last but not least, *researchers* use information from national journals for their daily scientific-scholarly activities.

*Concise review of informetric studies on a journal's national orientation*

Almost four decades ago Braun and Nagy (1982) carried out a comparative analysis of the nationality of authors publishing in Hungarian foreign language journals and in national journals from several other countries. Uzun (2004) analysed patterns in foreign authorship of articles and in the geographical origin of journal editorial board members in leading journals in the field of information science. Moed introduced an Index of National Orientation (INO), defined as the share of the papers from the country most frequently publishing in a journal, relative to the total number of papers published in the journal. (Moed, 2005).

Focusing on the measurement of national scientific publication productivity, Basu (2010) found that changes in a nation's productivity may be due to changes in the degree to which a database covers national journals from that nation. She addressed the meaningfulness of productivity measurement in terms of indexed papers in databases which change their journal coverage over time.

Considering Brazilian research output, Leta (2008) concluded that "Brazilian science is gaining space in international databases but it seems that some other requisites are needed to gain international audience, including the establishment of a strong national policy towards better training of researchers and journals in English proficiency" (p. 51). Still relating to Brazil, Vargas, Andrea de Souza Vanz & Stumpf, (2014) found that half of the Brazilian publication output in agriculture was published in national journals, many of which had only recently been included in the Web of Science, were written in Portuguese and had a low journal impact factor.

Analyzing a set of over 4,000 journals published by 3,500 national publishers during the time period 2000-2013, Gazni (2015) found that foreign authorship increased from 36 to 62 per cent during the period, but he observed large differences between countries and research disciplines. In an evaluation study of Polish journals Kulczycki and Rozkosz (2017) concluded that a multidimensional evaluation of local journals should not rely only on bibliometric indicators derived from WoS or Scopus, but must be complemented with expert-based assessments according to common guidelines. Kim, Kim & Kang (2018) reported an increase in citation impact measured in the SCI-Expanded (the key segment of Web of Science) of national Korean journals in engineering and natural sciences. Focusing on Spanish and Italian journals in social sciences and humanities, Aleixandre-Benavent et al. (2019) underlined that publication characteristics in these domains of scholarship differ strongly from those in other disciplines in terms of geographical scope of research, including its national or local orientation, document type, publication language and collaboration practices. Kulczycki et al. (2018) analysed patterns of journals in social sciences and humanities from 8 European countries.

In his essay on the relevance of national journals, Zheng (2019) argued that national academic journals may claim their value from the following four aspects. (i) At the academic resource level, they may cover the most important academic research in fields in which a country has a unique position faster, more comprehensively and systematically. Chinese journals covering traditional medicine are typical examples. (ii). At the methodological level, national journals may have a unique information organization and architecture. For instance, most Chinese journals include funding information as a stand-alone annotation in their manuscripts. (iii). As regards the efficiency of scholarly communication, national journals may be an integral part of a country's scientific and technical output. In large, non-English speaking countries this output may be much larger than that covered by the corpus of English-language journals. (iv). Papers published in national journals are a rapid and effective way for researchers to acknowledge and make visible the support from their national funding agencies. Furthermore, they enable the public supervision of academic ethics within a country.

*Scope of the current paper*

The current article builds further upon an earlier article published by Moed et al. (2020a; 2020b). A journal's national orientation is measured by two bibliometric indicators: the first is based on the affiliation countries of *publishing* authors, and the second on the affiliations of authors *citing* a particular journal. The key question addressed in the earlier as well as the current paper is: How does the geographical orientation and citation impact of nationally oriented journals entering Scopus in early years develop over time? Did national journals become more international and increase their impact as measured by citations?

Throughout this paper the term "national journal" is used to denote a journal whose publishing and citing *authors* work in institutions that are to certain degree concentrated in a single country. In the literature on the subject, this term is sometimes used to refer to the country of the *publisher*, but this use is problematic as many large publishers have offices or branches in multiple countries. Furthermore, it must be noted that an author's *affiliation* country may differ from his or her country of *nationality*, i.e., the country of which an author holds *citizenship.*

The current authors have created a new dataset extracted from SCImago Journal Rank that is richer and more up-to-date than the set used in the earlier study. Firstly, the objective of the current paper is to update and expand the most important analyses carried out in Moed et al. (2020a; 2020b). Secondly, it provides a critical methodological discussion of the validity of the two key bibliometric journal indicators, namely the Index of National Orientation (INO) and a field-normalized measure of citation impact. Thirdly, while the update of the earlier study focused on national journals in the Scopus database as a whole, the current paper proposes a methodology to analyse nationally oriented journals of a particular country, and compare these with national periodicals from other countries.

INO is defined as the share of the papers from the country most frequently publishing in (INO-P) or citing (INO-C) a journal, relative to the total number of papers published in the journal or citing it. A purely national journal would have an INO value of 100 per cent. It characterises a journal merely on the basis of information derived from its *own* articles. This indicator, however, has several limitations: it does not take into account international co-authorship or the extent to which a country is *over*represented as compared with its *total* publication output. In *Section 2*, alternative, normalized INO variants are explored and correlated with the original ones.

Field-normalized citation impact indicators are often used in informetric studies as they take into account differences in citation frequencies between scientific-scholarly subfields. *Section 3* addresses the following question: If one assesses the field-normalized impact of nationally oriented journals many of which have a relatively low impact factor, could a possible effect of their inclusion in the database be that they lower the average citation levels in the subject fields they cover, and make a field normalized impact indicator biased in favour of national journals?

An update and extension of the main research questions addressed in the earlier study (Moed et al., 2020a; 2020b) is presented in *Section 4*. Analysing a fixed cohort of national journals entering the database during 1997-2010 and following them for at least 10 years up until 2019, the current analysis dedicates more attention to citation impact, and analyses not only trends in indicators but also their actual values at the start and the end of the time period considered. Furthermore, it uses two INO thresholds for defining a nationally oriented journal (50 and 80 per cent), and systematically analyses five indicators: the number of articles published, their national orientation in terms of publishing and of citing author populations, and a straight as well as a field-normalized journal impact factor.

The outcomes of the trend analysis presented in Section 4 are affected by th statistical phenomenon of *regression toward the mean:* when a journal has a large INO-P value upon entrance in the database, its score will tend to return to the average later on. Applying a bordered symmetric random walk model, *Section 5* gives a rough indication of the extent to which this phenomenon affects the trend analysis in Section 4.

The outcomes presented in Section 4 are used as *benchmark* data in a study of national journals per country presented in *Section 6*. This section analyses national journals used by authors from a set of 11 countries including Russia, other former Soviet Republics, and former East European members of the USSR Pact, and denoted as "USSR-related" countries throughout this paper. In Section 5 this analysis is presented primarily as a methodological exercise. Its outcomes will be used in a future publication examining the relationship between author publication practices, national research assessment and funding criteria, and Scopus database coverage in USSR-related nations. Finally, Section 7 draws the main conclusions, and discusses a number of factors that one should take into account when interpreting indicators of national orientation.

*Important limitation*

There is a growing interest in studying and assessing national journals. Recently, Pölönen et al. (2020) published a comprehensive overview of the creation of what they term "national lists of journals" and their use in performance-based research funding systems. It must be underlined that the current paper focuses on a specific subset of national journals, namely those journals that show a strong national orientation in terms of the affiliation countries of publishing and citing authors, and, most importantly, that are *indexed in Scopus* and active in 2019.

*Current authors' view on the value of national journals*

National journals are valuable sources of scientific-scholarly information. The paper does not question this. Public policies for the evaluation of science have led a part of the scientific community to believe that national journals are of less scientific value than those denoted as international. The current authors do not share this vision. The current paper focuses on methodologies to study the possible effects of the inclusion of national journals in international databases. Journal internationality is *not* considered to be an *all-decisive norm* in scientific-scholarly publishing. As argued in Moed (2017): "Journals could be systematically categorized according to their function and target audience, and separate indicators could be calculated for each category. In an analysis of research output in journals directed towards national audiences, citation-based indicators are less relevant. At the same time, in citation analyses based on the large international citation indexes focusing on the international research front, it would be appropriate to disregard such journals. The question as to which weights should be given to the various aspects in an assessment should be answered in the assessment's evaluative framework" (Moed, 2017, p.117). The analysis of USSR-related countries presented in Section 6 does *not* aim to assess performance of countries, but to analyse differences among countries and to identify factors that may be responsible for these differences and should be studied in follow-up research.

## 2. Validity of a journal's Index of National Orientation (INO)

A journal's INO-P in a particular year is defined as the percentage share of articles published by authors affiliated with institutions located in the country that has contributed the *largest* number of articles published in that journal and year. Its value ranges between 0 and 100 per cent. For instance, a journal's INO value of 80 per cent means that there is one country that appears as author affiliation country in 80 per cent of the articles published in the journal. The indicator of national orientation used in the current article aims to characterize a given journal's distribution of articles among the affiliation countries of its authors. It focuses on the top of this distribution as it scans as it were the affiliation information of publishing authors in each paper, counts for each country the number of articles in which it appears at least once as author affiliation country, and identifies the country with the largest number of appearances. This indicator has the following limitations.

- *Multiple appearances* of a country in the affiliation list in a particular paper are *not* taken into account. A further refinement would be to sum up the number of *authorships* linked with a country in all papers published in a journal rather than the number of *articles*, or determining the number of *distinct* authors from a country, and identify the country with the largest share of authorships or authors. An alternative method is to sum up what can be termed as "*country-ships*", and determine the total number of times a particular country appears in a journal's combined author affiliation lists, including its multiple appearances in the same paper.

- INO does *not* take into account the phenomenon of *international co-authorship*. If a country would publish as the top country a stable share of articles in a national journal compared to that of other countries in the same journal, but increase its international collaboration and publish more and more internationally collaborative papers in the journal, this would not lead to an increase in the INO value, although if strong enough this trend would be visible when considering not only the 'top' country, but also the percentage of articles published by the next 3 or 5 countries. It must be noted that there are several separate indicators of international collaboration calculated and available at the level of journals as well (See Scimago Journal Rank).

- INO aims to characterise a journal purely on the basis of information derived from its *own* articles. It does *not* adopt the *perspective of a particular country* and assess the extent to which it is *as an affiliation country over*represented as compared to the total publication output it produces in the total set of journals indexed in the total database or in a particular subject field. For instance, a country publishing (more precise: contributing to) 10 percent of papers in a particular journal and one per cent of papers in the total database, publishes in the journal ten times as many papers as expected on the basis of its total publication output, while a second country publishing 10 per cent in the journal as well, but being more productive overall and publishing 10 per cent of papers in the database would have a number of papers in the journal "as expected". INO does *not* take into account such differences in overall productivity among countries.

The current section presents a technical discussion of this indicator and explores a more sophisticated but also more complex, normalized version of INO, taking into account the limitations mentioned above. In a first step, a new INO-like indicator was calculated based on a country's number of countryships in the sense defined above. In the calculation of a country's share of countryships the denominator is the total number of *countryships* over all countries publishing in a journal rather than the total number of *articles* published. A journal's INO based on countryships is then defined as the score of the country contributing the largest number of countryships to the journal. This approach can be conceived as a manner of calculating *fractional* publication counts, in which the total number of countryships plays the role of the "correct divisor" as defined by Narin et al. (1988).

A second approach does not only take into account the top country with the largest number of papers or countryships in a journal, but also all other affiliation countries. In addition, it accounts for the preference of each country to publish in the journal, defined as the ratio of the country's share of papers in the journal and its share of its papers in the entire database. A normalized indicator (NINO) could be defined as defined as follows:

$$NINO = \frac{\sum_{i=0}^{n} w_i \cdot p_i}{\sum_{i=0}^{n} w_i}, \ i=1…n$$

In which $p_i$ indicates a country's percentage of articles or countryships in a journal, $n$ the number of countries publishing in the journal and $w_i$ a weight factor. It assumes that the number of articles published by a country in the journal is the primary variable: the larger a country's share of papers in a journal, the larger its contribution to NINO. In this definition the value of *NINO* is expressed as a percentage in the range between 0 and 100, as is the case for the original *INO*. It is plausible to assume that the weight $w_i$ depends at least upon the following two key characteristics: the country's preference to publish in a journal, relative to the country's share of publications in the database (or in a particular subject field), and the importance of a country's contribution to the journal, relative to that of other countries publishing in the same journal.

In the approach adopted below the weight $w_i$ is defined as the product of two different operationalisations of these two characteristics.

$$w_i = p^k \cdot AI^m \ \text{k,m,>0}$$

In this formula *AI* is stands for Activity Index. *AI* is defined as the ratio of the percentage of a country's papers in a journal and its share of its papers in the entire database. If *AI* exceeds the value of 1.0, a country publishes more papers in the journal than expected on the basis of the country's share of papers in the total database. There are more sophisticated measures of preference than *AI,* but at least it is conceptually simple, is often applied in bibliometric studies, and can be used to examine a global effect upon the (normalized) INO value. *AI* is sensitive to outliers. Small countries with a low share of papers in the database, e.g., 0.1 per cent, but responsible for, say, 50 per cent of papers published in a particular journal, show an *AI* of 500 for that journal, while a large country with a world share of 10 per cent and account of 50 per cent of papers in another journal, have an *AI* of 5.0. To obtain at least some idea of the effect of such outliers, one *NINO* variant presented below in Table 1 is based on *AI itself* (i.e., exponent *m*=1) and a second one on the *square root* of AI rather than on *AI* itself (*m*=½).

The importance of a country's contribution to the journal is measured by its percentage of papers in the journal, *p*. It is true that this factor is also included as a parameter in the NINO formula above, but including it also in the weight factor makes it possible to vary its weight in the calculation. Table 1 presents a NINO variant in which no *additional* weight is given to *p* (exponent *k*=0), and one in which *k*=1, thus showing the effect of giving more weight to the share of a country's papers in a journal.

Table 1. Pearson correlation coefficients between six INO variants

|  | Original INO | INO based on country-ships | NINO weight= AI | NINO weight= sqrt(AI) | NINO weight= p*AI | NINO weight= p*sqrt(AI) |
|---|---|---|---|---|---|---|
| Original INO | 1.00 | 0.96 | 0.68 | 0.66 | 0.88 | 0.91 |
| INO based on countryships | 0.96 | 1.00 | 0.76 | 0.75 | 0.94 | 0.97 |
| NINO weight= AI | 0.68 | 0.76 | 1.00 | 0.98 | 0.86 | 0.85 |
| INO weight= sqrt(AI) | 0.66 | 0.75 | 0.98 | 1.00 | 0.82 | 0.82 |
| NINO weight= p*AI | 0.88 | 0.94 | 0.86 | 0.82 | 1.00 | 0.99 |
| NINO weight= p*sqrt(AI) | 0.91 | 0.97 | 0.85 | 0.82 | 0.99 | 1.00 |

Legend to Table 1. Data relate to the year 2019. All correlations are significant at p=0.001. Number of journals: analysed: 18,600. *AI*: Activity Index (see main text). *p*: Percentage of articles. Sqrt: square root.

Table 1 shows the Pearson correlations between 6 INO-related indicators calculated for all about 18,600 journals: the original INO, an INO variant based on countryships, and four normalized measures based on four definitions of the weight factor discussed above. Table 1 shows first of all that the original INO and the INO based on countryships strongly correlate; Pearson's R is as high as 0.96. Next, normalized indicators in which the weight factor includes only the Activity Index show the lowest correlation with the original INO. The dampening of AI values by calculating their square roots does not seem to have much effect in the set of journals as a whole. The indicators based on weight factors including a country's percentage of articles show a Pearson coefficient of 0.88 with the original INO, and 0.91 with the INO measure based on countryships.

This correlation analysis relates to the *total* set of journals. Journals used by large and productive countries such as USA and China tend to have somewhat lower NINO than INO values, and show lower Pearson correlation coefficients between original and normalized INO. For instance, for journals in which USA has the largest number of papers the correlation coefficient between original INO and NINO with a weight factor equal to *p*AI* amounts to 0.83 for USA, and for China and Russia to 0.90, while for the Netherlands it is 0.95 and for Ukraine 0.98. The fact that countries with large publication outputs such as USA reveal a somewhat weaker correlation with the original INO than small countries is as expected. But it must be underlined that the effect of the inclusion of *AI* in the weight factor is not merely *reducing* the contribution of countries with relatively *large* publication outputs, but also *amplifying* the role of *less* productive countries.

From a theoretical point of view, the issue rises which factor should be the most dominant when measuring a journal's national orientation: the distribution a journal's papers across countries, and especially the appearance of a highly productive country, or a country's overall level of publication productivity in the total database, and, if both factors are relevant, which one should prevail. This cannot be decided *purely* on statistical grounds, but requires a further theoretical clarification of the notion of national orientation, and of the context in which the indicator is to be used. The current authors do not claim that the NINO indicators explored above are the best possible measures in terms of statistical validity, nor that international co-authorship or a country's total volume of published articles are irrelevant aspects of a journal's international orientation. It would certainly be interesting to explore a normalized indicator of national orientation further, exploring more sophisticated alternatives of the Activitiy Index, but there are three limitations of such type of indicator.

*Firstly*, it is a *complex* indicator, and therefore rather difficult to explain to a non-bibliometric expert. The current, simple version can be easily calculated or checked from the online version of Scopus, which increases its transparency. Furthermore, it has a straightforward statistical interpretation. *Secondly*, the value of this normalized indicator is *database dependent*. If a database would be biased in favour of a particular country, the norm against which overrepresentation *within* a journal is evaluated may be biased itself: a country may be overrepresented *in the database*, compared to the country's share of active researchers in the world, or compared its share of papers in other databases. *Thirdly*, taking into account a country's total publication output embodies a shift in attention toward a possible *explanation* for the dominance of a particular author country in a journal. One may question the added value of including an explaining factor in an indicator itself, even though doing so is a common practice in the field of informetrics and bibliometrics, the target- or source-normalized journal impact factors being typical examples.

These three limitations lead the current authors to the conclusion that the original INO is a valid, fit-for-purpose, transparent statistical tool to identify journals with a strong national orientation, and to examine their role in the global scientific-scholarly communication system. Furthermore, these measures can be properly used to analyse trends in the journal coverage of a database, compare journal coverage between databases, and study publication practices of researchers and coverage policies of database producers. This is why the original INO is used as the preferred indicator in the current paper. In any case, the preliminary results presented above show that there is a very strong linear correlation between the original, simple INO and a complex, normalized version. This means that the first is a very good predictor of the second, and that, even if one believes that a normalised indicator as measure of national orientation is more valid than the simple version, the INO used in the current paper is still a good proxy of this concept.

The next sections examine trends in the national orientation and citation impact of journals which had a strong national orientation when they entered the database. A threshold value for INO-P of 50 per cent is chosen. The countries with the largest share of published articles in the database in the year 2019 are USA, China and Great Britain, with 17, 15 and 5 per cent, respectively. To reach an INO-P level above 50 per cent in a particular journal, the Activity Index for China and USA in that journal must exceed 3, which is still substantial, and that for Great Britain even a value of 10.

## 3. Validity of the field-normalised journal impact measure

A second indicator that must be discussed is a *relative or field-normalised impact* measure of journal J that is calculated by dividing J's 3-years impact factor by the average impact factor of all journals in the subject fields assigned to J. In this way, a value of 1.0 represents a 'world' subject field average. The journal impact factor calculated in the current study is similar to the 'official' JCR/Clarivate impact factor, but is based on citations in a particular year to articles published in the *three* rather than the *two* preceding years, and is denoted as a 3-year impact factor with symbol JIF.

The inclusion of journals with a relatively low citation impact in a database has an effect on its overall, "average" characteristics, especially on average citation rates in subject fields. Zitt and Bassecoulard have pointed towards this effect in several earlier articles (Zitt & Bassecoulard, 1998). A possible effect of the inclusion of low impact journals is that it lowers the average citation levels in the subject fields they cover. If one aims to assess the effects of inclusion of nationally oriented journals, many of which may be poorly cited when they enter the database, by comparing their citation impact with their subfield average and calculating a relative citation rate for them, these rates may be affected by the journals' inclusion itself. This could make the interpretation of trends in citation rates of national journals more difficult. The current section examines whether there is any effect at all, and what its order of magnitude would be.

If one makes a rough distinction between poorly cited and well cited journals in a particular subject field, and if the ratio of the number of poorly and well cited, active journals is constant over the years, there is a tendency that the mean or median JIF does not change. But if the ratio of the number of poorly and well cited journals *does* change over time, the mean (or median) JIF changes as well, but in the opposite direction. As a result, in this case the field-normalized JIF of journals entering the database in the beginning of the time period would increase if the ratio of poor to well cited journals grows over time, and decline if this ratio would go down. But this is true *both* for *poorly* cited *and* for *well* cited journals, and, expressed as a relative increase in the score in the end year compared to that in the begin year, both to the same degree. Therefore, when analysing poorly cited journals, the best solution to reduce the effect of changes in the ratio of poorly versus well cited journals is to compare the behaviour of poorly cited journals with that of their well cited counterparts.

The database analysed in the current paper continuously expands its coverage. The number of journals indexed in Scopus and still active in 2019 steadily more than doubled during 1996-2019, from 8,500 to more than 18,000 journals, while the median number of articles published in an indexed journal revealed during this time period a slight increase from around 45 to 50 articles per year. Furthermore, an increasing number of non-journal sources, mostly conference proceedings and books, were indexed as well. Therefore, it is no surprise that the impact factors of a fixed cohort of journals tend to increase during the time period considered.

Figure 1 shows that the ratio of number of national (INO-P>50) and non-national (INO-P<=50) journals and published articles declined during 1996-2019. Apparently, the contribution of national journals to the database declined over time, although the decline in the first years is partially due to the fact that during this interval and particularly between 2001 and 2003, for a considerable number of journals only affiliation data for the *first or corresponding author* was indexed. An additional analysis not presented in the current paper revealed that all disciplines show a declining trend in the ratio of the number of national to non-national journals and published papers, although for Clinical Medicine and especially for Social Sciences and Humanities it is much stronger than it is for the other fields. This ratio of median JIF of national to non-national journals declined in the first half of 1999-2019 but increased in the second half. All disciplines except one show this pattern. The only exception is Social Sciences and Humanities in which the ratio of the median impact factors of national versus non-national journals showed a decline over the *total* time period 1999-2019.

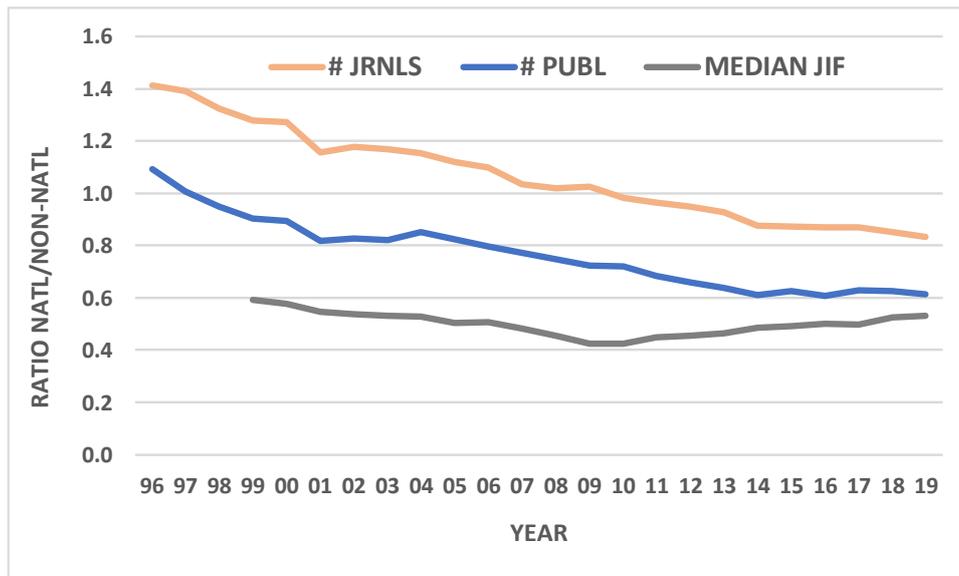

Figure 1. Ratio National (INO>50) /Non-national (INO<=50) number of journals, published articles and median JIF per year

As will be shown in Section 4, nationally oriented journals tend to have a substantially lower impact factor than more internationally oriented periodicals. Applying the considerations presented above as regards poorly versus well cited journals to the analysis of nationally versus internationally oriented journals, it is therefore not obvious to assume that when nationally oriented journals increase their relative citation impact – and Section 4 will show that there is a substantial fraction doing so – this increase is due to the fact that more national journals enter the database and therefore reduce the field averages in the database. If there is any effect of changes in the composition of journals in terms of national orientation in a subject field or the entire database, it works in the opposite direction. In other words, there is evidence for a slight negative bias towards national journals in the relative citation rates calculated in the earlier study and in the next section. The discipline most strongly affected by this negative bias against national journals is Social Sciences and Humanities.

If the ratio of national to non-national journals declines during the entire time period 1996-2019, and assuming that national journals have lower impact factors than their more nationally oriented counterparts, one would expect to a find a monotonously increasing ratio of the mean JIF over time rather than the kinked line displayed in Figure 1. It is hypothesised that this kink is due to the above mentioned fact that in the earlier years, especially during 2001-2003, the database producer indexed for a substantial number of journals only affiliation data of the first or reprint author. In fact, although small, Figure 1 reveals an *increase* in the ratio of national to non-national journals in 2001. It should also be noted that the increase in mean JIF in the second half of the time period is also partially due to the phenomenon that journals having a strong national orientation when they enter the database may internationalize and move during the time period considered from national to non-national journals and increase their citation impact. It is this factor that is further analysed in the next Section.

## 4. Characteristics of nationally oriented journals in Scopus

*Journal sets analysed in the current study*

The current paper analyses journals indexed in Scopus that meet the following criteria: they are *active* in 2019; they have an *uninterrupted* number of publications in Scopus in each year between the first publication year for which Scopus indexed their publications (shortly indicated as a journal's entry year or begin year in Scopus) up until and including 2019; journals entering Scopus after 2017 are *not* included. Next, a journal should publish on average at least 10 publications per year during its active time period. Publications include all documents categorized in Scopus as article, review, proceedings paper and short survey.

Not all documents indexed in Scopus contain affiliation information on publishing authors. In the calculation of the INO measures, publications without affiliations were *not* taken into account. The total number of articles used in INO's denominator does *not* include publications without author affiliations. If a journal's percentage of papers without affiliations was found to be above 50 per cent, it was excluded from the study set. However, in the calculation of the citation impact indicators publications without author affiliations were included. The total number of journal in the final study set amounts to 18,600.

*Trends in 5 key indicators of national journals entering Scopus during 1997-2010*

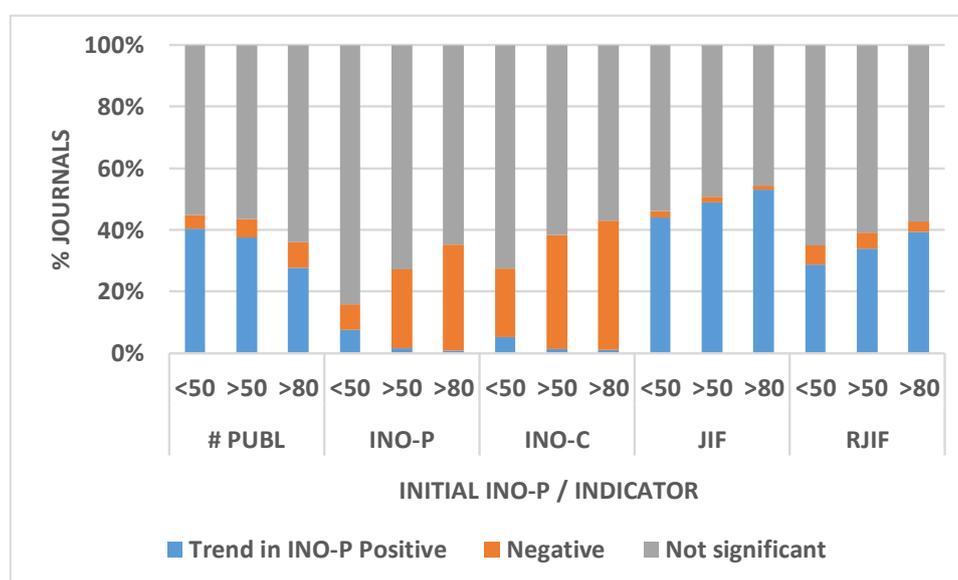

Figure 2: Trends in five indicators for three sets of journals entering Scopus between 1997-2010 and followed up until 2019. Initial INO-P <=50: n=2,400 journals; Initial INO-P>50: n=4,000; Initial INO-P>80: n=2,300. Significance level: p=0.01. # PUBL: Number of published articles; INO-P, INO-C: Index of National Orientation based upon the affiliations of publishing (P) or citing (C) authors; JIF: 3-year Journal Impact Factor; RJIF: Field-normalised (or Relative) Journal Impact Factor.

Figure 2 relates to journals entering the Scopus database between publication years 1997 and 2010 and displays for three sets of journals specified below the trend during the time period between their entry year and 2019 – i.e. during a time period of at least 10 years – of the following five key indicators: the annual number of articles published in a journal (# PUBL); The Indicator of a journal's National Orientation in terms of the affiliations of publishing (INO-P) or citing (INO-C) authors; and the 3-years Journal Impact Factor (JIF) as well as the Relative Journal Impact Factor, defined as the ratio of a journal's JIF value and the mean JIF value of all journals covering its subject field.

The three sets of journals analysed are: (i) About 2,400 journals entering Scopus during 1997-2010; with an initial INO-P value below 50 % in the year in which they entered Scopus. (ii) 4,000 Journals with an initial INO-P value above 50 %. (iii) 2,300 Journals showing a strong national orientation with initial INO-P values larger than 80 %. This set is a *subset* of the second. For each indicator a growth rate was computed over the years, based on the outcomes of a linear regression, with the indicator as the dependent and the year as independent variable, and by dividing the regression coefficient by the journal's mean annual score. It was tested whether the trend in annual scores was significant or not, applying a 99 per cent confidence level.

Focusing first on the set of journals entering the database during 1997-2010 with an initial value below *50 per cent*, Figure 2 shows that 40 per cent of journals shows a significant increase in the number of articles published annually. 8 % shows a significant decline in INO-P and 23 % in INO-C. 44 % of journals significantly increased their journal impact factor, but only 29 % showed significantly growing relative, field normalized impact factors.

Journals with an initial INO-P *above 50 per cent* reveal a larger share of periodicals with declining INO values (26 % for INO-P and 37 % for INO-C) and a larger share with increasing straight (49 %) or field-normalised (34%) journal impact factors than journals with an initial INO-P below 50 % do. Journals with INO-P above 80 % show the same tendency as those with initial INO-P above 50 %. This suggests that the statistical relationship between the number of journals showing a significant decline trend in national orientation or increase in journal impact on the one hand, and the threshold value for initial national orientation on the other, is *monotonous*. In the analyses presented in the next chapter this threshold is set to 50. But the outcomes suggest that the observe tendencies would become even stronger if this threshold is raised.

*Trends* reveal only a part of the statistical tendencies. The *levels* of the values of the key indicators in the begin and end year are relevant as well. Objects with extreme values at the start can be expected to show a significant trend more often than objects having average start values. Therefore, Table 2 presents information on the levels of the indicators. As in Figure 2, Table 2 relates to journals that were indexed for the first year in Scopus during 1997-2010 and gives statistics for each of the three sets of journals analysed in Figure 2.

Table 2 presents for the five key indicators their *median* values in the first year a journal was indexed in Scopus and in the last year (2019), as well as the ratio of the medians in these two years. In all cases the standard deviation exceeds the mean, which is a common phenomenon in bibliometrics and reflects the skewness of the distributions studied. This is why medians rather than averages were calculated. Focusing on the subset of journals with an initial INO-P above 50 per cent, Figure 2 reveals that the median INO value for publishing authors declined from 86 to 63 per cent, and that for citing authors from 80 to 42 %. As regards citation impact, more than half of the journals increased their 3-year impact factors, raising the median value from around 0.2 to 0.9, and almost 40 percent increased their field-normalized impact, the median value of which grew from 0.2 to 0.5.

Table 2. Median values of 5 indicators in begin and end year for the three journal sets

| INO-P threshold | Indicator | Begin Year | End Year | Ratio Score End Yr / Begin Yr | Indicator | Begin Year | End Year | Ratio Score End Yr / Begin Yr |
|---|---|---|---|---|---|---|---|---|
| 0-50 | # PUBL | 24 | 44 | 1.8 | | | | |
| 50-100 | | 26 | 43 | 1.7 | | | | |
| 80-100 | | 29 | 45 | 1.6 | | | | |
| 0-50 | INO-P | 36 | 33 | 0.9 | INO-C | 43 | 29 | 0.7 |
| 50-100 | | 86 | 63 | 0.7 | | 80 | 42 | 0.5 |
| 80-100 | | 96 | 78 | 0.8 | | 100 | 50 | 0.5 |
| 0-50 | JIF | 0.5 | 1.5 | 2.9 | RJIF | 0.4 | 0.8 | 1.9 |
| 50-100 | | 0.2 | 0.9 | 4.3 | | 0.2 | 0.5 | 2.8 |
| 80-100 | | 0.2 | 0.7 | 4.8 | | 0.1 | 0.4 | 3.1 |

*The role of research discipline, publication language and access status*

Table 3 presents results per main research discipline, publication language, and access status. Comparing 5 main disciplines, Social Sciences and Humanities shows the lowest median number of published articles and the largest INO-C value in 2019, i.e., the strongest national orientation in terms of citing author affiliations. Clinical Medicine has the largest INO-P, and Biomedical Research the largest median number of publications, JIF and field-normalised JIF.

Differences among publication languages are large. To give one example, for journals publishing in non-English languages the median INO-P value in 2019 is 89 per cent, against 55 per cent for periodicals publishing only in English. Only 17 per cent of non-English publishing journals showed a significant decline in their INO-P value, and 26 % in INO-C. For journals publishing in English these percentages are almost twice as high: 35 and 44 %.

Table 3. Results per main discipline, publication language and OA status

| Factor | # Journals | % Jrnls with significant increase up until 2019 in: | | | % Jrnls with significant decline in: | | Score in 2019 | | | | |
|---|---|---|---|---|---|---|---|---|---|---|---|
| | | # PUBL | JIF | RJIF | INO-P | INO-C | # PUBL | JIF | RJIF | INO-P | INO-C |
| *Main discipline* | | | | | | | | | | | |
| Social Sci & Humanities | 1,384 | 26 % | 45 % | 31 % | 24 % | 36 % | 27 | 0.7 | 0.5 | 62 | 45 |
| Clinical Med | 1,447 | 36 % | 55 % | 43 % | 37 % | 47 % | 58 | 1.1 | 0.5 | 67 | 39 |
| Biomed Res | 535 | 36 % | 48 % | 40 % | 41 % | 54 % | 64 | 2.0 | 0.7 | 57 | 38 |
| Natural Sci | 1,025 | 32 % | 56 % | 37 % | 32 % | 38 % | 53 | 1.0 | 0.5 | 61 | 44 |
| Engineering | 709 | 36 % | 57 % | 36 % | 30 % | 34 % | 59 | 1.1 | 0.5 | 56 | 41 |
| *Publication Languange* | | | | | | | | | | | |
| English | 2,794 | 37 % | 54 % | 39 % | 35 % | 44 % | 46 | 1.3 | 0.7 | 55 | 37 |
| English+Other | 422 | 14 % | 46 % | 34 % | 27 % | 36 % | 34 | 0.4 | 0.3 | 78 | 55 |
| Non-English | 6,96 | 21 % | 44 % | 32 % | 17 % | 26 % | 43 | 0.3 | 0.2 | 89 | 74 |
| *Access status* | | | | | | | | | | | |
| In DOAJ/ROAD | 1,227 | 32 % | 56 % | 44 % | 36 % | 43 % | 46 | 0.9 | 0.5 | 62 | 38 |
| not in DOAJ/ROAD | 2,772 | 32 % | 49 % | 34 % | 28 % | 38 % | 42 | 0.9 | 0.5 | 64 | 44 |

For journals included in DOAJ/ROAD according to information in the Scopus Source Journal List (Scopus, 2020) the percentages of journals showing a significant decline in INO-P or INO-C are larger than those calculated for non-DOAJ/ROAD journals (36 vs. 28 % and 43 vs. 38 %, respectively). But the initial values when they entered the database were somewhat higher as well (88 vs. 84 for INO-P and 83 vs. 79 % for INO-C). In 2019 the median INO-P is only slightly lower than that for non-OA journals (62 vs. 64 %), but the median INO-C is substantially lower (38 against 44 %), while the field-normalised citation impact is the same for both (RJIF=0.5). This outcome provides evidence that OA journals in 2019 tend to be somewhat broader in terms of geographical coverage of publishing and citing authors. but their citation impact is not necessarily larger than that of non-OA periodicals.

## 5. A statistical note

In Moed et al. (2020) the authors underline that the outcome of the trend analysis of journals entering the database with a strong national orientation is partially a statistical artefact. This is due to *regression toward the mean*, i.e., the statistical phenomenon that arises if a sample point of a random variable is extreme, a future point will be closer to the mean or average on further measurements. When a journal has a large INO-P value upon entrance in the database, its score will tend to return to the average later on. This phenomenon complicates the interpretation of the outcomes of the trend analysis, as one would expect purely on statistical grounds that a part of the journals showing a significant decline in INO-P would reveal such a decline anyway, regardless of their degree of integration in the international journal network. But how large is this part? Is it possible to give at least a rough indication of the extent to which this phenomenon affects the trend analysis?

To illustrate the statistical artefact, Moed et al. (2020) gave the following example: if one selects from a set of objects with randomly assigned binary scores (0 or 1) in two subsequent years a sub-set of those objects that have score 1 in the first year, the probability that the score of journals in this sub-set is in the second year lower than that in the first year (i.c., score 0) is 50 per cent. Building upon this example, the base idea below is to derive an indication of the role of chance from an analysis of temporal patterns in the scores of journals. What is the probability that the observed patterns found in these journals are generated by chance?

It analyses journals entering the database between 1997 and 2010 with INO-P values above 90 per cent. The total number of these journals amounts to 1,645. 758 of these show a significant negative trend according to a linear regression outlined in Section 4. The number of years that a journal can be followed obviously depends upon the entry year in the database, and ranges between 10 (for journals indexed for the first time in 2010) to 23 years (for sources entering in 1997). INO-P values are arranged into 10 classes (0-10%, 10-20%, etc.) numbered 1 to 10. For instance, score 10 relates to INO-P values between 90 and 100 per cent. For each journal the difference is calculated between the score obtained in the end year and that in the begin year, and is denoted below as the journal's *net decline* during the time period considered.

A simple bordered symmetric random walk model was applied, in which one moves forward from left to right in a two dimensional lattice with 10 rows (representing scores) and 23 columns (representing years). The transition probabilities from one state to another depend on the location of the state as on margin and corner states the movement is limited. For non-margin or non-corner states the following three transitions have the same probability: moving diagonally upward to the right (increase, symbol '+'), moving horizontally one step to the right (remaining constant, 'o') and moving diagonally downwards to the right (decline, '-'). For corner or margin states, the probability to move outside the lattice is zero.

Starting the walk from the upper left node, representing the state of a journal in its entry year with INO-P score 10, two paths are allowed: either constancy ('o') or decline ('-'). This case represents the example mentioned above. Adding a second step generates 5 paths: oo, o-,-+,-o and --. In two cases (oo,-+) the score obtained at the end of the path equals the start value. The net decline is zero. For two more cases (o-,-o) it is one, while in one case (--) it amounts to 2. The latter result can be interpreted as follows: a journal entering with the largest score 10 in its entry year has after two years a probability of 20 per cent (one out of five) to reveal a score of 8, i.e., a net decline of 2. This process can be continued by adding more steps.

Elementary calculus enables one to calculate after each step the probability that a journal "choosing" in each subsequent year at random between decline, increase or remaining constant, has revealed a particular net decline compared to its initial score. As an example, Figure 3 presents the outcomes of this model calculation for the case that the number of steps equals 17. It also gives the distribution of net decline values in 2019 among all 126 journals entering the database in 2002, and for the subset of 48 journals showing a significant decline according to the linear regression analysis.

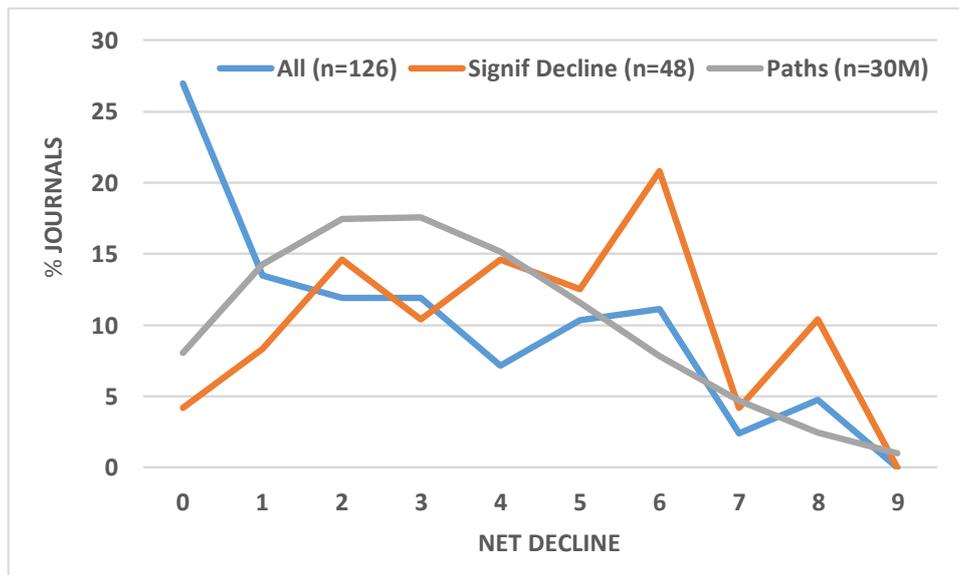

Figure 3. Percentage of journals as a function of the net decline (=difference between score in end year and begin year). All: Set of all 126 journals active in 2019 with entry year 2002 and INO-P>90%; Signif Decline: The subset of 48 journals revealing a significant decline according to the test based on linear regression (Section 4). Paths: Based on the bordered symmetric random walk model (see main text).

Figure 3 shows for the total set of 126 journals that the percentage of journals showing a zero net decline is much higher than that based on the random walk model. Focusing on the set of journals with a significant decline according to the linear regression analysis, Figure 3 shows that, compared to the outcomes of the random walk model, journals with low net decline values tend to be underrepresented and those with large net declines overrepresented. The results for other entry years show similar patterns, although the amount of variability in those related to the set of significantly declining journals is large, due to a small number of these journals (for some years less than 25).

The fact that journals with a zero net decline are so strongly overrepresented in the set of *all studied journals* compared to the distribution of net declines generated by the random walk model suggests that structural factors are working on this set of strongly nationally oriented journals that prevents them from internationalizing.

On the other hand, the observation that the distribution of net decline values among journals in the set of *journals with a significant decline* differs from this distribution predicted by the random walk model and biased in favour of high decline values, provides evidence that the formation of this subset cannot be ascribed merely to chance, and at least partially reflects a genuine statistical tendency to show a decline in terms of national orientation.

These results do not allow one to further quantify this tendency. A more detailed analysis should also take into account journals with lower INO-P values, for instance, between 50 and 90 per cent.

Moreover, it must be noted that the analysis presented in this section is fully based on discretisation of percentages into 10 intervals. Mathematically, it is possible to make these intervals smaller and redo all calculations.

## 6. Results on former USSR-related countries

*Publications in Russian national journals*

While the previous sections analysed the total set of journals entering the database during 1997-2010, the current section relates to the subset of nationally oriented journals with INO-P above 50 % from former USSR republics and East European countries under the influence for the former USSR. In the current section, a national journal is defined as a journal in which there is one author affiliation country accounting for at least 50 per cent of all articles published in the journal. An additional distinction is made between *domestic* and *foreign* national journals. From the point of view of a particular country C, if authors affiliated with institutions located in C publish in a journal nationally oriented towards country D, the journal is said to be *domestic national* if C and D are the same, and *foreign national* if they are different.

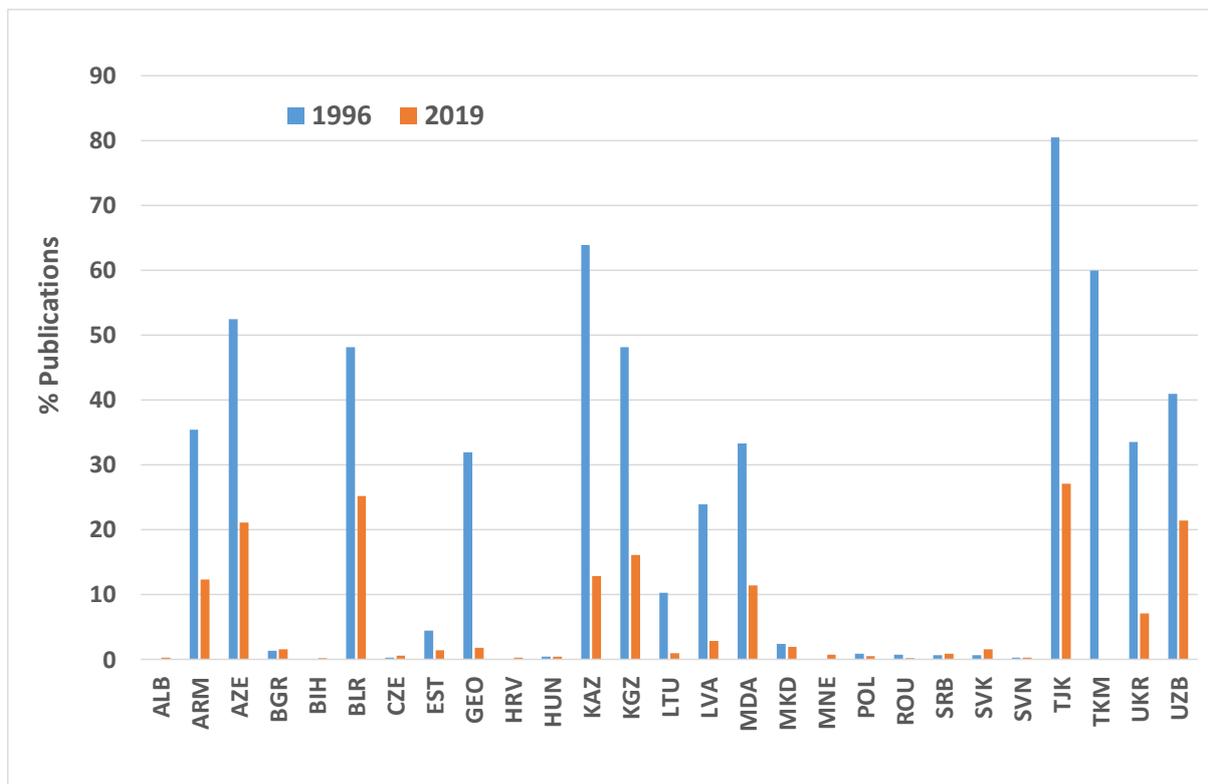

Figure 4. Percentage of publications in national Russian journals (INO-P>50% in 2019) in 1996 and 2019. Large difference exist in publication counts between countries. Publication counts for 2019: Above 10,000: Russia, Poland, Czech Republic. Below 1,000: Albania, Bosnia & Herzegovina, Moldova, North Macedonia, Montenegro, Tajikistan, Turkmenistan, Uzbekistan. All other countries have between 1,000 and 10,000 publications.

Figure 4 deals with one specific type of papers in foreign national journals, namely articles in *Russian* national journals. For all countries that published to a certain degree in Russian journals this share has strongly declined. These are all former USSR republics, while former communist Central and Eastern European countries hardly published in Russian journals, nor in the beginning nor in the end of the time period, except to some extent the Baltic States.

*Trends in national journals*

Figures 5 and 6 below present outcomes for the 11 countries having more than 5 national journals entering Scopus during 1997-2010. They shed light on the question as to what extent these journals have become more international in terms of publishing author population and citation impact. As presented in Figure 2 in Section 4, for the total set of nationally oriented national journals in Scopus the percentage of journals with a significant decline in INO-P is 26 per cent. Using this percentage as a reference, it can be concluded from Figure 5 that Lithuania and Serbia are substantially above this reference level, with 56 % and 53 %, respectively, while for Russia, Ukraine and Hungary it is below this level, with values of 11, 15 and 17 %, respectively.

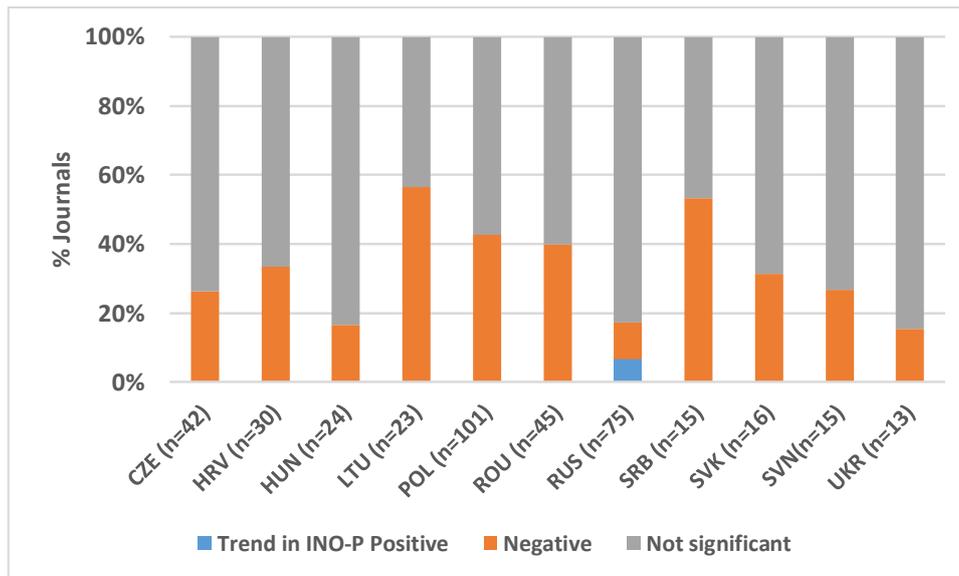

Figure 5. Trends in Indicator of National Orientation (INO-P) of the national journals in which a country has published

Considering field-normalised citation impact in Figure 6, Russia and Poland show the largest percentage of national periodicals that manage to increase their relative citation impact, namely 71 and 50 percent, respectively. This is substantially above the level of 34 per cent that relates to all national Scopus journals and that is displayed in Figure 2. For Romania and Serbia, the percentage is near this level, while for the other countries it is below it.

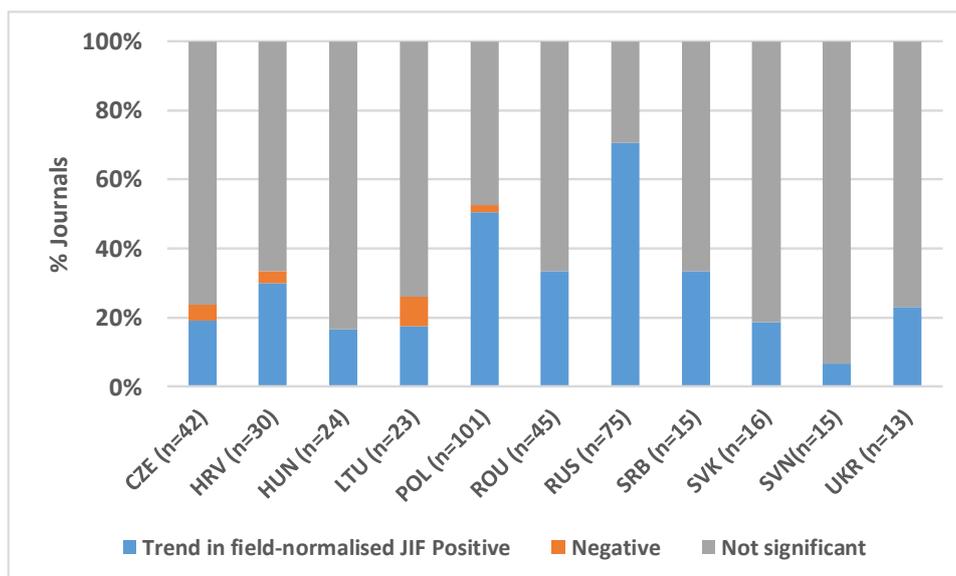

Figure 6. Trends in Field-Normalised Impact of national journals in which a country has published

*Level of scores in begin and end year*

While Figures 5 and 6 provide insight into the statistical significance of *trends* in two key indicators calculated for a country's national journals entering Scopus during 1997-2010, Figures 7-9 give information about the *level* of the scores in begin and end year. Figure 7 shows that for most countries the median number of publications in the selected set of national journals either decreased or increased only slightly. The striking exception is Russia as its annual number of publications in national journals increased from 50 to more than 70, in both years much larger than the median values for the total database presented in Table 2 in Section 4 (26 and 43 in begin and end year, respectively).

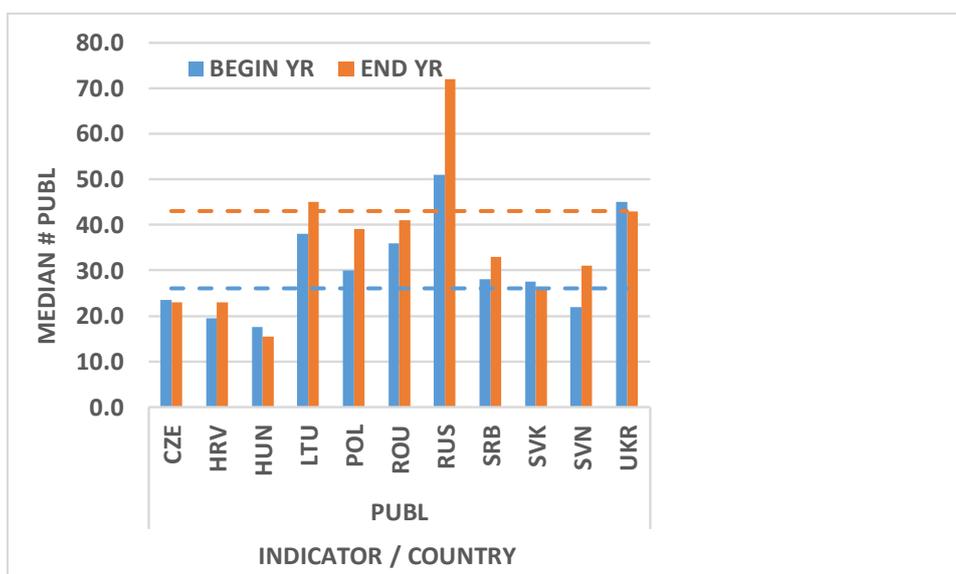

Figure 7. Median number of publications in begin and end year in national journals (INO-P>50%). The dashed lines give the median values for the total database presented in Table 2 in Section 4.

The median INO-P and INO-C values of national journals in begin and end year are plotted in Figure 8. Using the median INO-P value of 86 % in the begin year and 63 % in the end year calculated for all national journals in Scopus and presented in Table 2 as reference values. Figure 8 shows that in the begin year the majority of selected countries have an INO-P value above this level, but in the end year all but three are below the reference for that year. For INO-C the decline in the median INO-C values in the last year compared to the begin year is eve**n** more pronounced than it is for INO-P. In the begin year all 11 countries except one have an INO-C value *above* the reference value of 80.0 for all Scopus journals, while in the end year 7 have a value *below* the reference value of 42.0 %. This outcome provides an indication that both the author population and the citation impact the major part of the selected countries' national journals tend to become more internationally oriented during the time period considered.

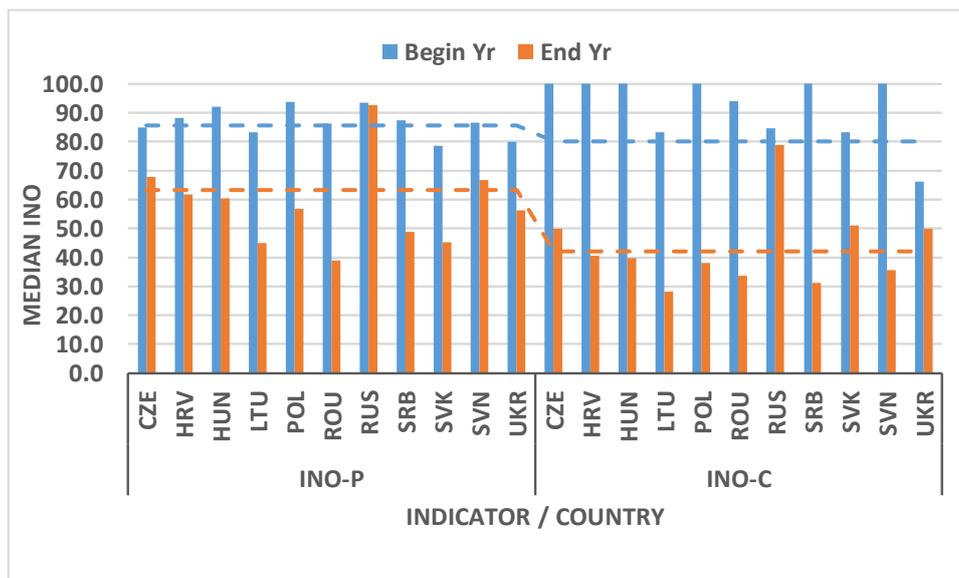

Figure 8. Median indicator of national orientation of national journals in begin and end year. INO-P: national orientation as expressed in country affiliations of the publishing authors; INO-C: as INO-P, but relating to citing rather than publishing authors. The dashed lines give the median values for the total database presented in Table 2 in Section 4.

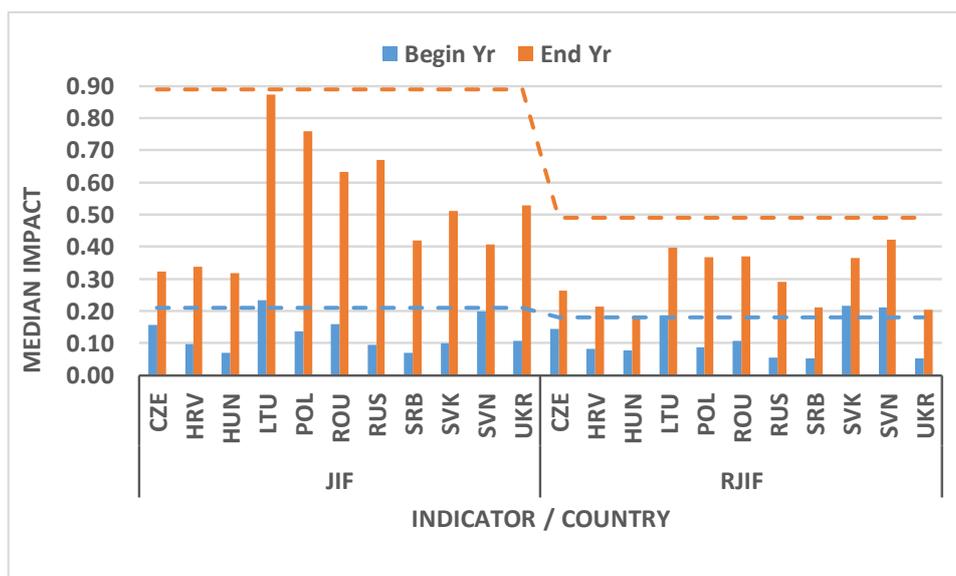

Figure 9. Median citation impact of national journals in begin and end year. JIF: 3-year Journal Impact Factor. RIF: Relative (field-normalized) Journal (1.0= field average). The dashed lines give the median values for the total database presented in Table 2 in Section 4.

Figure 9 relates to the citation impact of the national journals. For all countries except one the median journal impact factor of national journals is in the begin year *below* the Scopus reference value of 0.2, and in the end year none of the journals exceeds the median level calculated for all Scopus journals in that year. As journal impact factors are affected by differences in subject fields, it is more informative to consider field-normalized impact measures. But the outcomes are hardly different: in the begin year only three and in the end year none of the selected journals have a field-normalized impact exceeding the median of 0.5 calculated for all Scopus journals in those years. It must be concluded that the citation impact of the countries' national journals, though having increased over time, tends to be still relative low compared to other national journals in the field and indexed in Scopus.

## 7. Discussion and conclusions

*General conclusions*

The current article has shown that the relatively simple measure of national orientation used in the current and in earlier papers, aiming to characterise national orientation as a percentage of papers published by the most productive country in a journal, uses only data obtained from the journal itself, and strongly correlates with more sophisticated indicators that take into account international collaboration and a country's share of articles in the total base, or, alternatively, in the subject field the journal covers. Although the current authors do not claim that the sophisticated indicators explored in the paper are the best possible measures from a statistical-informetric point of view, or that international collaboration and country productivity are per se irrelevant aspects, the INO measures can be considered good proxies of national orientation, and are easy to explain and to verify in an online database.

The main conclusions drawn in the earlier paper are confirmed in the current study: Focusing on journals with INO-P above 50 per cent, 26 per cent of nationally oriented journals entering Scopus during 1997-2010 show a significant decline in the national orientation of their publishing author populations, and 37 per cent reveal a decline in that of their citing author populations. These percentages are higher than those obtained for the set of more internationally oriented periodicals with initial INO below 50 %. The median INO value for publishing authors declined from 86 to 63 %, and that for citing authors from 80 to 42 %.

As regards citation impact, more than half of the journals increased their 3-year impact factors, raising the median value from around 0.2 to 0.9, and almost 40 percent increased their field-normalized impact, the median value of which grew from 0.2 to 0.5. These percentages of journals revealing a significant increase in journal impact are *larger* than those in the set of more internationally oriented periodicals, despite the fact that the methodological discussion in Section 3 identified a negative bias against nationally oriented journals. Selecting in the study set journals with INO-P above 80 rather than 50 percent revealed the same tendencies, but somewhat sharper. This provides evidence that the conclusions drawn on nationally oriented journals do not depend upon the initial INO threshold value that was chosen to define a national journal.

The outcomes provide evidence that nationally oriented journals do not constitute a separate, isolated segment of Scopus. Instead, many national journals become more integrated in the global journal network after they entered the database. Perhaps the most informative result revealing this tendency is the strong decline in the journals' median INO-C, the percentage of citations from the most frequently citing affiliation country. This measure is not affected by an overall increase in international scientific collaboration. Combined with the observed increase in citation impact, it follows that for many journals the citation impact does not merely increase in size, but also broadens in terms of geographical scope. This observed tendency is in agreement with the hypothesis of Acharya et al. (2014) stating that the impact of non-elite journals is growing.

*Factors to be considered*

It was found that *publication language* and *Open Access Status* have a positive effect upon the tendency to internationalize in terms of declining INO values and increasing citation impact measures. The percentage of journals showing a significant decline in national orientation using non-English publication language is only 17 per cent, against 35 per cent for periodicals with English as publication language. However, more research is needed as to which factors are responsible for the differences between journals from the same country and subject field. The current sub-section discusses factors related to scientific information retrieval, international scientific migration and collaboration, database coverage policy, the size of a national research community, political-historical factors, and national research assessment and funding policies.

Authors publishing in nationally oriented journals can themselves increase the probability of being retrieved in a citation index, by choosing informative titles and author-given keywords, study research topics sharing a wide interest, and, last but not least, cite key articles dealing with the same topics. Perhaps these *information retrieval-related characteristics* provide a key to the question as to which additional factors may cause or reinforce differences in the degree to which national journals become more internally oriented in terms of publishing and citing authors.

Furthermore, it must be noted that an author's *affiliation* country is not the same as his or her country of *nationality*. For instance, when a scientist from country C moves abroad (semi-) permanently and publishes in one of C's national journals or cites one in his or her articles published in internationally oriented journals, this is not counted as a national publication or citation. *Scientific migration* and *international collaboration* is increasing and can be expected to have an effect on the position of national journals in the international journal network as well.

*Database coverage policy* is an important factor influencing the degree to which a database covers nationally orientated journals. Scopus and Web of Science have different coverage policies. The latter indexes less nationally oriented journals than the first, and gives more weight to citation frequency as selection criterion. The implication is that the outcomes obtained in the current *case study* based on *Scopus* may differ from results obtained in an analysis of Web of Science or another multi-disciplinary citation index.

The number of national journals covered in Scopus reveals large differences *between countries*. For instance, Figure 4 in Moed et al. (2020b) shows that more than 25 per cent of articles by authors affiliated with institutions in Russia, China, Brazil and Ukraine in 2014-2016 were published in national journals (with INO-P>80%). For USA and Great Britain these percentages were 12 and 5, respectively. The analysis for the USSR-related countries in Section 4 in the current paper revealed large differences among countries both regarding the extent to which they published in nationally oriented journals indexed in Scopus, as well as in the trends in the indices of national orientation and citation impact of their national journals. These differences may be due to specific database coverage policies and market strategies of the database producer.

The *size of a national research community* plays a role as well, as national journals are economically viable only in countries with a large academic work-force. Furthermore, the analyses of USSR-related countries also clearly illustrated the importance of *historical-political factors* in the publication practices of author, although it must be noted that the fact that the former USSR republic shared Russian as a common language is an important factor as well.

Criteria applied in *public academic policies* at a national level should be considered as well. Hladchenko and Moed (n.d.) analysed policies in Ukraine for the attainment of a doctorate or for the promotion to associate professor and professor. These policies did *not* prioritise publications in reputable peer-reviewed journals with an international orientation, which resulted in an *increase* in the number of publications in *national* Ukrainian periodicals. Future studies must further unravel the relationships between author publication practices in a country, the criteria used in national research assessment and funding policies, and database coverage policies.